\documentclass[aps,pra,twocolumn,showpacs,superscriptaddress,groupedaddress]{revtex4-1}  

\usepackage{graphicx}  
\usepackage{dcolumn}   
\usepackage{bm}        
\usepackage{amssymb}   
\usepackage{amsmath}

\usepackage[colorlinks=true,breaklinks=true,allcolors=blue]{hyperref}
\usepackage{url}
\usepackage{txfonts}

\usepackage{graphics}
\usepackage{pstricks}
\usepackage{tikz}
\usepackage{graphicx}
\usepackage{color}
\usepackage{hyperref}

\newcommand{\eq}[1]{(\ref{eq:#1})}
\newcommand{\Eq}[1]{Eq.\,\eqref{eq:#1}}
\newcommand{\Fig}[1]{Fig.~\ref{fig:#1}}

\begin{document}


\title{Bi-directional universal dynamics in a spinor Bose gas close to a non-thermal fixed point}
\author{Christian-Marcel Schmied}
\author{Maximilian Pr\"ufer}
\author{Markus K. Oberthaler}
\author{Thomas Gasenzer}
\affiliation{Kirchhoff-Institut f\"ur Physik,
             Ruprecht-Karls-Universit\"at Heidelberg,
             Im~Neuenheimer~Feld~227,
             69120~Heidelberg, Germany}
\date{\today}

\begin{abstract}
We numerically study the universal scaling dynamics of an isolated one-dimensional ferromagnetic spin-1 Bose gas. 
Preparing the system in a far-from-equilibrium initial state, simultaneous coarsening and refining is found to enable and characterize the approach to a non-thermal fixed point. A macroscopic length scale which scales in time according to $L_{\Lambda}(t)\sim t^{\, \beta}$, with $\beta\simeq 1/4$, quantifies the coarsening of the size of spin textures.
At the same time kink-like defects populating these textures undergo a refining process measured by a shrinking microscopic length scale $L_{\lambda}\sim t^{\, \beta'}$, with $\beta'\simeq-0.17$.
The combination of these scaling evolutions enables particle and energy conservation in the isolated system and constitutes a bi-directional transport in momentum space. 
The value of the coarsening exponent $\beta$ suggests the dynamics to belong to the universality class of diffusive coarsening of the one-dimensional XY-model.
However, the universal momentum distribution function exhibiting non-linear transport marks the distinction between diffusive coarsening and the approach of a non-thermal fixed point in the isolated system considered here.
This underlines the importance of the universal scaling function in classifying non-thermal fixed points.
Present-day experiments with quantum gases are expected to have access to the predicted bi-directional scaling.
\end{abstract}


\maketitle

\begin{figure*}
\includegraphics[width=0.83\textwidth]{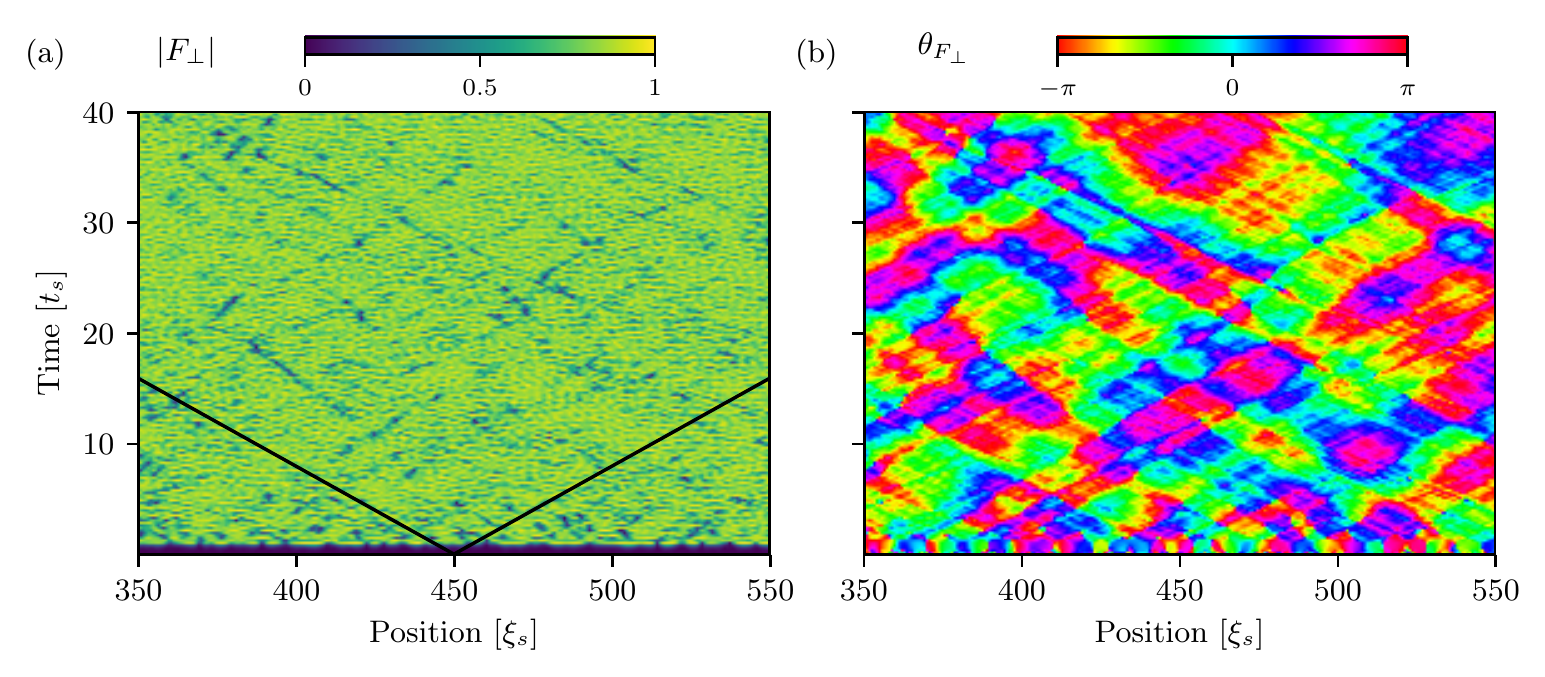}
\caption{\label{fig:RealSpaceEvoSpin} 
Space-time evolution of the transversal spin $F_{\perp} = \lvert F_{\perp}\rvert \exp \left \{{i \theta_{F_{\perp}}}\right \}$. 
Quenching the system across the quantum phase transition, here to $q_\mathrm{f}=0.9$, introduces exponentially growing unstable modes which subsequently lead to the formation of textures in the transversal spin after a few characteristic time scales $t_\mathrm{s}$. 
These spin textures, with size given by the distance over which a $2 \pi$ phase winding occurs in the phase angle $\theta_{F_{\perp}}$, are populated by kink-like defects.  
The defects are characterized by a dip in the amplitude and a corresponding phase jump as depicted in panels (a) and (b). The solid black lines in panel (a) indicate a sound cone associated with the sound velocity of the spin degree of freedom $c_\mathrm{s}= \sqrt {n_0 \lvert c_1 \rvert  } = 1$.
The size of the spin textures grows in time which is associated with the dilution of kink-like defects leading to long-range order developing in the phase field (see panel (b)). 
Each panel only shows an excerpt of the total grid of length $\mathcal{L}= 554\,\xi_\mathrm{s}$.
}
\end{figure*}

\section{Introduction}

The dynamics of isolated quantum many-body systems quenched far out of equilibrium has been studied extensively in the recent past.
Nonetheless, many questions remain open concerning, in particular, possible universal scaling on the way to equilibrium. 
For example, during prethermalization \cite{Gring2011a,Langen2015b.Science348.207,Aarts2000a.PhysRevD.63.025012,Berges:2004ce,Langen:2016vdb} quasi-stationary mode occupancies show trivial scaling in time, $\sim t^{0}$.  
Further universal phenomena include many-body localization \cite{Schreiber2015a.Science349.842}, critical and prethermal dynamics \cite{Braun2014a.arXiv1403.7199B,Nicklas:2015gwa,Navon2015a.Science.347.167N,Eigen2018a.arXiv180509802E,Smale2018a.arXiv180611044S}, decoherence and revivals \cite{Rauer2017a.arXiv170508231R.Science360.307},  as well as wave- and superfluid turbulence \cite{Zakharov1992a,Nazarenko2011a,Navon2016a.Nature.539.72,Navon2018a.arXiv180707564N,Gauthier2018a.arXiv180106951G,Johnstone2018a.arXiv180106952}. 
Beyond these scenarios, if scaling occurs simultaneously in time and space, the evolution can become a type of renormalization-group flow, with time as the flow parameter, associated with the existence of a non-thermal fixed point \cite{Berges:2008wm,Berges:2008sr,Scheppach:2009wu,Prufer:2018hto,Erne:2018gmz}.
Such fixed points have been discussed and experimentally observed with \cite{Nowak:2010tm,Nowak:2011sk,Schmidt:2012kw,Schole:2012kt,Karl:2013mn,Karl:2013kua,Karl2017b.NJP19.093014,Erne:2018gmz}  and without \cite{Berges:2008wm,Berges:2008sr,Scheppach:2009wu,Berges:2010ez,Berges:2013fga,Orioli:2015dxa,Berges:2015kfa,PineiroOrioli:2018hst,Prufer:2018hto} reference to ordering patterns and kinetics, and topological defects, paving the way to a unifying description of universal dynamics.


Universal scaling in time and space of correlations of macroscopic observables in isolated systems is associated with the loss of information about the details of the initial condition and microscopic system properties.
This universal scaling evolution is closely related to transport in momentum space associated with a few relevant symmetries only, and the corresponding conservation laws \cite{Orioli:2015dxa,Chantesana:2018qsb,Schmied:2018upn,Mikheev2018a.arXiv180710228M}.
The ensuing universality renders such dynamics of strong interest across many different fields, including, besides cold gases, early-universe cosmology \cite{Kofman:1994rk, Micha2003a,Berges:2008wm,Gasenzer:2011by}, and quark-gluon dynamics induced by nuclear collisions \cite{Berges:2013fga,Berges:2014bba}.

In open classical systems coupled to a bath, universal scaling appears commonly in the context of dynamical critical phenomena \cite{Hohenberg1977a,Janssen1979a}, coarsening and phase-ordering kinetics \cite{Bray1994a.AdvPhys.43.357}, as well as glassy dynamics and ageing \cite{Calabrese2005a.JPA38.05.R133}.
For many-body quantum systems prethermal scaling \cite{%
DallaTorre2013.PhysRevLett.110.090404,
Gambassi2011a.EPL95.6,
Sciolla2013a.PhysRevB.88.201110, 
Smacchia2015a.PhysRevB.91.205136,
Maraga2015a.PhysRevE.92.042151,
Maraga2016b.PhysRevB.94.245122,
Chiocchetta2015a.PhysRevB.91.220302,
Chiocchetta2016a.PhysRevB.94.134311,
Chiocchetta:2016waa.PhysRevB.94.174301, 
Chiocchetta2016b.161202419C.PhysRevLett.118.135701} 
as well as coarsening dynamics \cite{%
Damle1996a.PhysRevA.54.5037,
Mukerjee2007a.PhysRevB.76.104519,
Williamson2016a.PhysRevLett.116.025301,
Hofmann2014PhRvL.113i5702H,
Williamson2016a.PhysRevA.94.023608,
Bourges2016a.arXiv161108922B.PhysRevA.95.023616}
has been studied.

Coarsening is a specific type of universal scaling evolution, generically associated with the phase-ordering kinetics of a system coupled to a temperature bath, and exhibiting an ordering phase transition \cite{Bray1994a.AdvPhys.43.357}.
Typically, the coarsening evolution following a quench into the ordered phase only involves a single characteristic length which fixes the scale of the correlations and evolves as a power law in time.
Such a growing length scale can be associated with, e.g., the coarsening of magnetic domains in the ordered phase.

In general, universal scaling manifests itself in the evolution of correlations described, e.g., in momentum space, by a structure factor $S(\mathbf{k},t)$ obeying the scaling form
\begin{equation}
  S(\mathbf{k},t)=\left(t/t_{\mathrm{ref}}\right)^{\alpha}f_\mathrm{s}\left( \left[t/t_{\mathrm{ref}}\right]^{\, \beta}\mathbf{k}\right)\,,
  \label{eq:Scaling}
\end{equation}  
where $f_\mathrm{s}$ is a universal scaling function depending on a single variable only. The corresponding scaling exponents $\alpha$ and $\beta$ define the evolution of the single characteristic length $L(t)\sim t^{\,\beta}$.
The time scale $t_{\mathrm{ref}}$ denotes some reference time within the temporal scaling regime.

The concept of non-thermal fixed points generalizes such universal scaling dynamics to isolated systems far from equilibrium.
Scaling dynamics near such a fixed point is associated with the transport of a conserved quantity, implying a behaviour \eq{Scaling} within a region of momenta in which the integral $\int\mathrm{d}\mathbf{k}\,k^{\alpha/\beta-1}S(k,t)$, with $k=|\mathbf{k}|$, remains invariant. 
Different conserved quantities can emerge in different momentum regimes, and thus the scaling dynamics violates single-length scaling.
In this case the scaling evolution is expected to be characterized by multiple length scales $L_{i}(t)$ with, in general, different scaling exponents.

Here we numerically demonstrate the violation of single-length scaling dynamics in one spatial dimension by studying the time evolution of an isolated spinor Bose gas after a sudden quench into the magnetically ordered phase.
We find a bi-directional self-similar evolution of the structure factor characterized by the algebraic growth of an infrared (IR) scale $L_{\Lambda}(t) \sim t^{\, \beta}$ associated with the conservation of local spin fluctuations as well as an algebraic decrease of a second scale $L_{\lambda}(t) \sim t^{\,\beta^{\prime}}$ connected to kinetic energy conservation in the ultraviolet (UV).
The growth of $L_{\Lambda}(t)$ is observed to be associated with the dilution of kink-like defects separating patches of approximately uniform spin orientation, 
while $L_{\lambda}(t)$ is set by the decreasing microscopic width of the defects, cf.~Ref.~\cite{Schmidt:2012kw}. 
Constraining the system to one spatial dimension, we find the IR scaling exponent $\beta\simeq0.25$.
This value is considerably smaller than the standard exponent $\beta=1/2$  found in isolated systems for universal scaling transport towards the IR \cite{Schole:2012kt,Orioli:2015dxa,Berges:2015kfa,Karl2017b.NJP19.093014,Chantesana:2018qsb,Prufer:2018hto,Schmied:2018upn,Mikheev2018a.arXiv180710228M}, associated with near-Gaussian fixed points \cite{Karl2017b.NJP19.093014,Mikheev2018a.arXiv180710228M}, and, in open systems in two and three dimensions, for diffusive coarsening of a non-conserved order parameter field \cite{Bray1994a.AdvPhys.43.357}. 
We emphasize that our findings are also in contrast to the case of a one-dimensional (1D) single-component gas where no scaling evolution is expected due to kinematic constraints on elastic $2 \to 2$ scattering from energy and particle-number conservation and $\beta\simeq0.1$ has been observed experimentally \cite{Erne:2018gmz}.

\section{Spin-1 Bose gas in one spatial dimension} 
\label{sec:Hamiltonian}

We consider a homogeneous one-dimensional spin-1 Bose gas described by the Hamiltonian \cite{Stamper-Kurn2013a.RevModPhys.85.1191}
\begin{equation}
H = \int \mathrm{d}x \left [ \vec{\Phi}^{\dagger} \left( -\frac{\hbar^{2}}{2M} \frac {\partial^2}{\partial x^2} + q f_z^2 \right) \vec{\Phi}+  \frac{c_0}2  n^2 + \frac{c_1}2  \lvert  \vec{F} \rvert ^2 \right],
\label{eq:Hamiltonian}
\end{equation}
where $\vec{\Phi} = \left(\Phi_1, \Phi_0, \Phi_{-1} \right)^T$ is a three-component bosonic spinor field whose components account for the magnetic sublevels $m_F = 0, \pm 1$ of the $F=1$ hyperfine manifold. 
$q$ is the quadratic Zeeman energy shift which is proportional to an external magnetic field along the $z$-direction. 
It leads to an effective detuning of the $m_F=\pm 1$ components with respect to the $m_F = 0$ component. We are working in a frame where a homogeneous linear Zeeman shift has been absorbed into the definition of the fields. 
Spin-independent contact interactions are described by the term $c_0 n^2$, where $n = \vec{\Phi}^{\dagger} \vec{\Phi}\equiv \sum_{m}{\Phi}_{m}^{\dagger} {\Phi}_{m}$ is the total density. 
Spin-dependent interactions are characterized by the term $c_1 \lvert \vec{F} \rvert^2$, where $\vec{F} = \vec{\Phi}^{\dagger} \vec{f}\, \vec{\Phi}$ is the spin density and $\vec{f} =  (f_x, f_y,f_z )$ is the spin in the fundamental representation. 
This term accounts. among others, for the redistribution of atoms between the three hyperfine levels \cite{Stamper-Kurn2013a.RevModPhys.85.1191}.
\begin{figure*}
\includegraphics[width=0.9\textwidth]{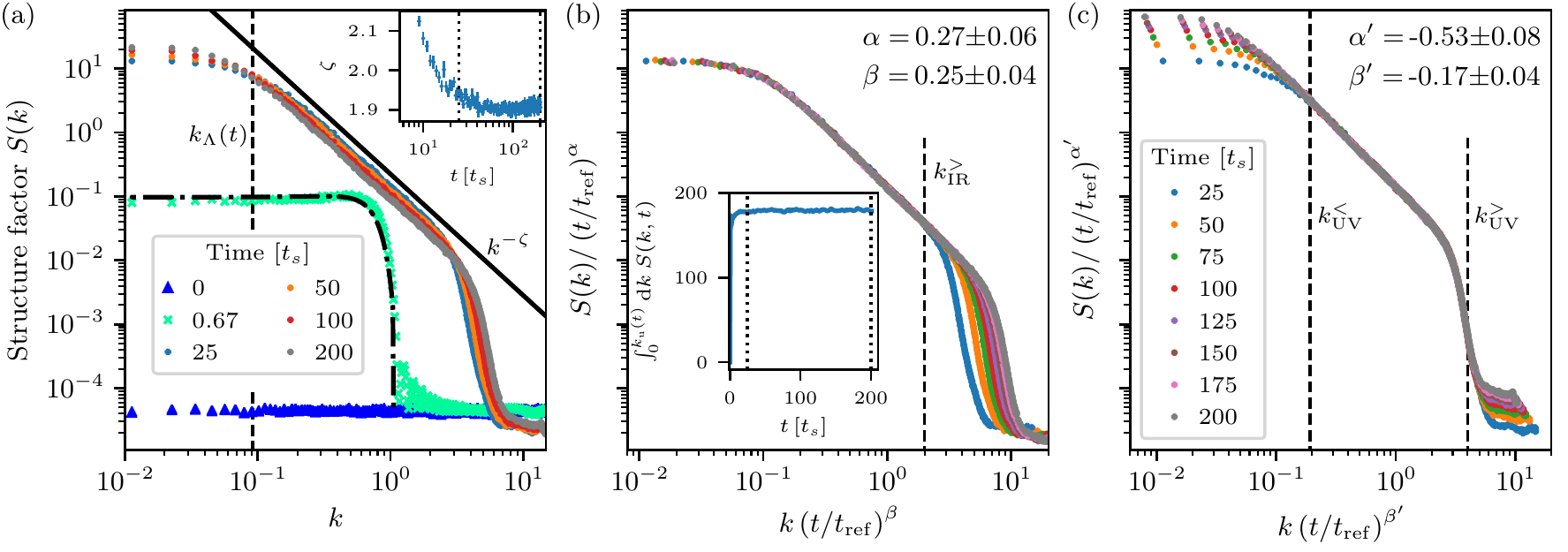}
\caption{\label{fig:StructureFactor}
Two-scale self-similar evolution of the structure factor $S(k,t)$. 
(a) Overview of the time evolution. 
The initial polar condensate at $t=0$ shows ground-state fluctuations around zero spin (dark blue triangles). 
At $t=0.67\, t_\mathrm{s}$ (green crosses) the population of the momentum modes is well approximated by the Bogoliubov prediction (dash-dotted line). 
For times $25\, t_\mathrm{s} \leq  t \leq 200\, t_\mathrm{s}$ the system is in the spatio-temporal scaling regime and evolves in a self-similar manner (dots). 
Three qualitatively different momentum regimes emerge: 
A plateau below a characteristic momentum scale $k_{\Lambda}(t)$ (dashed line exemplarily marks the scale at time $t = 25 \, t_{\mathrm{s}}$), a $k^{-\zeta}$ power-law fall-off at momenta up to a scale $k_{\lambda}$(t) and a steeper power-law decay at large momenta.
The inset shows the extracted exponent $\zeta$ (cf.~solid line in main frame).
(b) Structure factor in the temporal scaling regime rescaled according to \Eq{Scaling} with scaling exponents $\alpha = 0.27 \pm 0.06$ and $\beta= 0.25\pm0.04$ extracted via a least-square fit. 
Within the infrared scaling regime, $k\leq k_\mathrm{IR}^{>}$ (dashed line), all curves collapse onto a single one. 
The inset shows that, within this momentum regime, the local spin fluctuations are conserved (up to $\sim 2 \%$) for times $25\, t_\mathrm{s} \leq  t \leq 200\, t_\mathrm{s}$ (dotted lines).
The corresponding upper bound for the integral is set by $k_{\mathrm{u}}(t) = k_\mathrm{IR}^{>} \cdot (t/t_{\mathrm{ref}})^{- 1/4}$ with $t_{\mathrm{ref}} = 25 \, t_{\mathrm{s}}$.
Within errors, the scaling exponents are consistent with  $\alpha =  \beta$. 
(c) Structure factor in the temporal scaling regime rescaled according to \Eq{Scaling} with $\alpha^{\prime} = -0.53 \pm 0.08$ and $\beta^{\prime}= -0.17 \pm 0.04$. 
Within the ultraviolet scaling regime,  $k_\mathrm{UV}^{<} \leq k\leq k_\mathrm{UV}^{>}$ (marked by dashed lines), all curves collapse onto a single one. 
}
\end{figure*}

Spinor Bose gases can be realized in experiment in a well-controlled manner which makes them suitable for studying non-equilibrium phenomena \cite{Stenger1999a, Ho1998a,Stamper-Kurn2013a.RevModPhys.85.1191}, see Refs.~\cite{Sadler2006a, Bookjans2011a,Kawaguchi2012a.PhyRep.520.253,Prufer:2018hto} for dynamics after a quench.
Recently, universal scaling dynamics close to a non-thermal fixed point, with scaling exponent $\beta \simeq 1/2$, has been observed experimentally in a ferromagnetic ($c_{1}<0$) spin-1 system in a near-1D geometry \cite{Prufer:2018hto}. 
Theoretically, phase-ordering dynamics and scaling evolution has been studied in a ferromagnetic spin-1 Bose gas in 2D \cite{Williamson2016a.PhysRevLett.116.025301,  Williamson2016a.PhysRevA.94.023608,  Williamson2017a.PhysRevLett.119.255301, Symes2017a} as well as 
in a ferromagnetic spin-1 Bose-Hubbard model in a 1D optical lattice \cite{Fujimoto2018a}.

Apart from the trapping potential and a larger total density, we here perform  numerical simulations in the parameter regime realized in the experiment  \cite{Prufer:2018hto} on $^{87}$Rb in the $F=1$ hyperfine manifold. 
Assuming a constant homogeneous mean density $n_{0}=\langle n\rangle$, 
we can express the Hamiltonian \eq{Hamiltonian} in terms of the dimensionless length $\tilde x=x/\xi_\mathrm{s}$, with spin healing length $\xi_\mathrm{s} = \hbar/ \sqrt{2 M n_0 \lvert c_1 \vert} $, time $\tilde t=t/\tau_\mathrm{s}$, with spin-changing collision time $\tau_\mathrm{s}=t_\mathrm{s}/(2\pi)=\hbar/(n_{0}|c_{1}|)$.
The quadratic Zeeman shift is quantified by the dimensionless field strength $\tilde q=q\tau_\mathrm{s}/\hbar$, the field operators become $\tilde \Phi_{m}=\Phi_{m}/\sqrt{n_{0}}$, the density $\tilde n=n/n_{0}$, the spin vector $\tilde{\mathbf{F}}=\vec F/n_{0}$, and the dimensionless couplings read $\tilde c_{0}=c_{0}/|c_{1}|$ and $\tilde c_{1}=c_{1}/|c_{1}|=\mathrm{sgn}(c_{1})$. 
In the following, all quantities are expressed in the above units and the tilde will be suppressed.

In the ferromagnetic case ($c_1 = -1$),
and for a positive quadratic Zeeman energy $q$, the equilibrium system exhibits two different phases separated by a quantum phase transition that breaks the $U(3)$ spin symmetry of the ground state \cite{Kawaguchi2012a.PhyRep.520.253}.  
For $q> 2$ the system, in its mean-field ground state, is in the polar phase and thus unmagnetized.
On the opposite side of the transition, $0 < q <  2$, 
the ground state is in the easy-plane ferromagnetic phase.
Here, the non-conserved two-component order parameter is the transversal spin $F_{\perp} = F_x + i F_y$.
Hence, the mean spin vector is lying in $F_{x}$--$F_{y}$-plane, bearing magnetization $ \lvert F_{\perp}\rvert =  (1- q^{2}/4 )^{1/2}$. 

\section{Universal scaling dynamics} 
\label{sec:Results}
\subsection{Initial conditions and quench} 
\label{sec:NumericalResults}

We consider far-from-equilibrium dynamics after a quench, exerted on a homogeneous condensate in the polar phase, i.e.~an initial state with $\phi_{0}(x)=\langle\Phi_{0}(x)\rangle\equiv1$, by means of a sudden change of the quadratic Zeeman shift to the parameter range $0<q_\mathrm{f}<2$.
We compute the time evolution of observables using truncated Wigner  simulations, starting each run with a field configuration for  $q_\mathrm{i} \gg 2 $, with additional quantum noise added to the Bogoliubov modes  \cite{Blakie2008a, Polkovnikov2010a} of the polar condensate 
(see the appendix for details).

The quench induces transversal spin modes in the system to become unstable, leading to the formation of a spin-wave pattern  during the early-time evolution after the quench.
Non-linear interactions subsequently give rise to the formation of patches in the transversal spin.
Within each patch, the phase angle of the complex order parameter $F_{\perp}= \lvert F_{\perp} \rvert \exp \{i \theta_{F_{\perp}}\}$ is approximately constant in space (see \Fig{RealSpaceEvoSpin}b).
At the same time, defects, represented by a dip in the amplitude and a corresponding phase jump, are traveling across the system at roughly the speed $c_\mathrm{s}= \sqrt {n_0 \lvert c_1 \rvert  } = 1$ associated with the sound velocity of the spin degree of freedom (see solid lines in \Fig{RealSpaceEvoSpin}a).
Spin patches in combination with phase jumps form spin textures whose size is given by the distance over which a $2 \pi$ phase winding occurs.
The so-formed spin structure sets the stage for the subsequent ordering process.
According to the evolution charts in \Fig{RealSpaceEvoSpin} the average size of the textures 
appears to grow in time.

\subsection{Scaling evolution} 
\label{sec:ScalingEvolution}

For a quantitative analysis of the observed phase-ordering dynamics we consider averaged correlations of the order-parameter field.
Since our system is translationally invariant on average, we evaluate these correlations in momentum space, by means of the  structure factor
\begin{equation}
\label{eq:StructureFactor}
S(k,t) = \langle \lvert F_{\perp} (k,t ) \rvert^2 \rangle, 
\end{equation}
$\langle \dots \rangle$ denoting the average over different runs. 
$S(k,t)$ is formally obtained as the Fourier transform of $C(r,t)=\langle F_{\perp}(x,t)F_{\perp}(x',t)\rangle$ with respect to $r=x'-x$.

\Fig{StructureFactor}a shows the time evolution of the structure factor $S(k,t )$ for a quench to $q_\mathrm{f} = 0.9$ in the easy-plane ferromagnetic phase. 
The polar condensate at $t=0$ has no magnetization. 
At $t=0.67\, t_\mathrm{s}$ the population of the momentum modes of the structure factor within the instability regime fits the Bogoliubov prediction given by
$S (k,t) =4  \sinh^2 \left( \gamma_k t \right) /{ \gamma_k^2}$.
Here, the growth rate of unstable momentum modes $\gamma_k =\sqrt {\left( \epsilon_k+ q \right) \left(2- \epsilon_k -q  \right)}$, with mode energy $\epsilon_k = k^2$, is obtained as the imaginary part of the complex Bogoliubov mode energy \cite{Kawaguchi2012a.PhyRep.520.253}.
In the course of the subsequent non-linear redistribution of the excitations the system is found to enter a spatio-temporal scaling regime where the structure factor evolves in a self-similar manner. 
During this period of the relaxation process, we observe three qualitatively different momentum regions which reflect the patterns seen in the single spatial realizations in \Fig{RealSpaceEvoSpin}.
Below a characteristic momentum scale $k_{\Lambda}(t)$, the structure factor $S(k,t)$ shows a plateau.
Kink-like defects account for the power-law fall-off of the structure factor $S(k,t) \sim k^{-\zeta}$ for momenta $k\lesssim k_{\lambda}(t)$ \cite{Bray1994a.AdvPhys.43.357}. 
The exponent $\zeta=d+n$ depends on the dimensionality of the system and the defect structure. 
For kink-like defects ($n=1$) in one spatial dimension, the resulting exponent $\zeta=2$ is close to the value $\zeta = 1.91 \pm 0.02 $ which we extract by fitting the scaling form $A/[1+(k/k_{\Lambda})^{\,\zeta}]$ to the IR part of $S(k,t)$.
For large momenta $k\gtrsim k_{\lambda}(t)$, the structure factor shows a steeper fall-off, before saturating at the level of ground-state fluctuations, $S(k\to k_{a})\simeq0.5\times10^{-4}$, where $k_{a}$ denotes the lattice cutoff.

\Fig{StructureFactor}a indicates that the structure factor exhibits scaling according to \Eq{Scaling} within a region of IR momenta below the UV end of the $k^{-\zeta}$ power-law fall-off, i.e., for $k\lesssim k_{\lambda}(t)$. 
Taking the structure factor  at time $t_{\text{ref}} = 25\,t_\mathrm{s}$ as a reference and performing a least-square fit of the data up to $t = 200\, t_\mathrm{s}$ yields 
$\alpha =0.27 \pm 0.06$ and 
$\beta =0.25 \pm 0.04$.
The errors are determined from the width of a Gaussian distribution used to fit the marginal-likelihood functions of both scaling exponents \cite{ Orioli:2015dxa}. 
These errors can become relatively large due to statistical uncertainties and systematic deviations caused by the limited scaling window. 

Rescaling the structure factor in time by making use of the scaling form  \eq{Scaling} yields the collapse onto a single curve below the momentum scale $k_\mathrm{IR}^{>}$, as shown in \Fig{StructureFactor}b. 
The inset in \Fig{StructureFactor}b demonstrates that the local spin fluctuations are conserved in time (up to a relative error of $2 \%$)
within the IR scaling regime.
Hence, we find, to a good approximation, that $\partial_t \int_{0}^{k_{\mathrm{u}}(t)} \mathrm{d}k \, S (k,t) = 0$, with $k_{\mathrm{u}}(t) = k_\mathrm{IR}^{>} \cdot (t/t_{\mathrm{ref}})^{- 1/4}$ and $t_{\mathrm{ref}} = 25 \, t_{\mathrm{s}}$. Using the scaling form \eq{Scaling} for the structure factor $S(k,t)$ results in the scaling relation $\alpha = \beta$. 
The numerically extracted exponents are consistent with this scaling relation. 
As a consequence of the conserved local spin fluctuations we can describe the time evolution of the system for momenta $k\lesssim k_{\lambda}(t)$ by a single scaling exponent and thus a single characteristic IR length $L_{\Lambda}(t)\sim t^{\,\beta}$ with $\beta \simeq 0.25$. 
This macroscopic length scale corresponds to the size of the spin textures in the system.

\subsection{Violation of single-length scaling} 
\label{sec:ViolationSingleLS}

In the UV range of momenta, however, the structure factor violates this single-length scaling and rather suggests a second characteristic length  $L_{\lambda} (t)$ which shrinks in time.
\Fig{StructureFactor}c shows that the rescaled structure factor collapses onto a single curve for momenta $k_{\Lambda}(t) \lesssim k\lesssim k_\mathrm{UV}^{>}$ when choosing the scaling exponents $\alpha^{\prime} =-0.53 \pm 0.08$ and $\beta^{\prime} =-0.17 \pm 0.04$. 

Hence, we find that the structure factor, in a range of momenta with strong spin-wave excitations, obeys the extended scaling form \cite{Chantesana:2018qsb}
\begin{equation}
\label{eq:twoLScalingForm}
S( k,t ) = L_{\Lambda}(t)^{\alpha/\beta} f_{\mathrm{s}}(L_{\Lambda}(t)k, L_{\Lambda}(t)/L_{\lambda}(t)),
\end{equation}
with the scaling function being well approximated by
\begin{equation}
\label{eq:twoLScalingFunction}
f_{\mathrm{s}} (x,y) = f_{0}[1+x^{\,\zeta}+x^{\,\zeta^{\prime}}y^{\,\zeta-\zeta^\prime}]^{-1}\,.
\end{equation}
Here $\zeta^\prime \simeq 12\gg\zeta$ characterizes the large-$k$ fall-off.
For the bi-directional scaling behavior shown in \Fig{StructureFactor}, the scaling function $f_\mathrm{s}$ follows, to a good approximation, the form \eq{twoLScalingFunction}, with a single power-law exponent $\zeta$ in between the IR and UV scales $k_{\Lambda}(t)$ and $k_{\lambda}(t)$, respectively.
For the scaling function \eq{twoLScalingFunction}, the temporal scaling evolution of $L_{\Lambda}(t)$, $L_{\lambda}(t)$ implies that the exponents are related by 
$\alpha - \alpha^{\prime} = (\beta-\beta^{\prime}) \zeta\,$.
Moreover, imposing kinetic energy conservation in the UV scaling regime,  i.e.~$\partial_t \int_{k_{\mathrm{l}}(t)}^{k_{\mathrm{u}}(t)} \mathrm{d}k \, k^z \, S (k,t) = 0$ within a corresponding UV momentum interval $[k_{\mathrm{l}}(t)$, $k_{\mathrm{u}}(t)]$, yields $\alpha^{\prime} = \left (1+z\right) \beta^{\prime}$.
Taking the additional conservation of local spin fluctuations in the IR, $\alpha = \beta$, one obtains the dynamical exponent $z$, characterizing the dispersion $\omega (k)\sim k^z$, to be
\begin{equation}
z = \left ( \zeta -1 \right) \left(1- \beta/\beta^{\prime} \right).
\end{equation} 
Inserting the extracted parameters $\beta = 0.25 \pm 0.04$, $\beta^{\prime} = -0.17 \pm 0.04$ and $\zeta = 1.91 \pm 0.02$, we find
$z = 2.24 \pm 0.38$. 
To cross-check this result we numerically determine the dispersion for which the kinetic energy $\int_{k_{\mathrm{l}}(t)}^{k_{\mathrm{l}}(t)} \mathrm{d}k \, \omega (k) \, S (k,t)$ shows the minimal deviation from being conserved within the UV scaling regime set by $k_{\mathrm{l}}(t) = k_\mathrm{UV}^{<} \cdot (t/t_{\mathrm{ref}})^{0.17}$ and $k_{\mathrm{u}}(t) = k_\mathrm{UV}^{>} \cdot (t/t_{\mathrm{ref}})^{0.17}$ with $t_{\mathrm{ref}} = 25 \, t_{\mathrm{s}}$.
This method yields $z = 2.15 \pm 0.05$ consistent with the dynamical exponent directly calculated from the extracted scaling parameters.

\begin{figure}
\includegraphics[width=0.96\columnwidth]{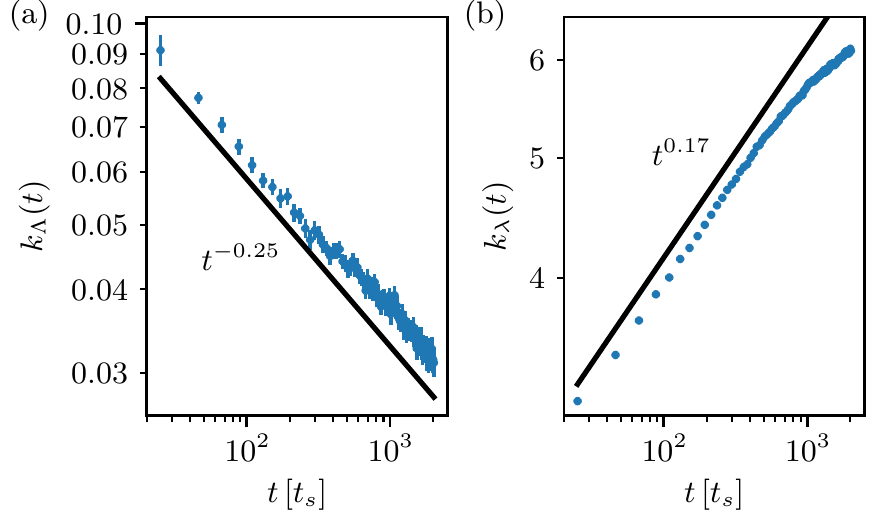}
\caption{\label{fig:Flowplot}
Time evolution of the characteristic IR scale $k_{\Lambda}(t)$ (panel (a)), as well as the UV scale $k_\lambda(t)$ (panel (b)).
While the IR scale decreases algebraically as $k_{\Lambda}(t)\sim L_{\Lambda}(t)^{-1} \sim t^{\, -0.25}$ over the whole range of evolution times considered (see solid line in (a)), the UV scale starts to deviate from  $k_{\lambda}(t)\sim L_{\lambda}(t)^{-1} \sim t^{\, 0.17}$ (see solid line in (b)) at $t \simeq 500 \, t_\mathrm{s}$. The deviation arises as the corresponding UV length scale approaches the spin healing length, i.e.~as $L_\lambda (t) = 2\pi/k_\lambda (t) \to 1$. 
While this causes the system to leave the regime of two-scale universal dynamics, it stays close to the non-thermal fixed point as the IR scaling behavior remains unchanged. 
The characteristic scales $k_{\Lambda}(t)$ and $k_\lambda(t)$ are obtained by means of fitting the scaling form \eq{twoLScalingFunction} to the structure factor up to a maximum momentum given by $k_{\mathrm{UV}}^{>}$ (see \Fig{StructureFactor}c).
We remark that the scaling function, \Eq{twoLScalingFunction}, does not appropriately capture the data in the UV regime of momenta anymore for evolution times $t \gtrsim 2000 \, t_\mathrm{s}$, which strongly influences the extraction of the UV scale $k_\lambda (t)$ (cf.~\Fig{StructureFactorUltraLong} showing $S(k,t)$ for evolution times up to $t= 4000 \, t_\mathrm{s}$).
Hence, we restrict our analysis to the time window $25 \, t_\mathrm{s} \leq t \leq 2000 \, t_\mathrm{s}$. 
Errors bars correspond to the fit error of the extracted scales. Note the double-log scale.
}
\end{figure}

\section{Discussion and conclusions} \label{sec:Conclusion}

As the scaling dynamics of the spinor system takes place in the transversal spin, it is instructive to compare with similar behavior known for the 1D XY model.
Considering an open system coupled to a heat bath and applying a temperature quench into the ordered phase leads to phase-ordering kinetics with a temporal scaling exponent $\beta = 1/4$ \cite{Rutenberg1995a.PhysRevLett.74.3836}.
At first sight, this appears to provide the universality classification for the self-similar dynamics seen in our system.
However, the nature of the respective evolutions turns out to be qualitatively very different.
Coarsening in the 1D XY model with non-conserved order parameter can be described as free phase diffusion of the order-parameter phase angle. 
As the topological charge is locally conserved in the system, the position-space correlation function at large distances $r$ is given by $C(r,t) \rightarrow \exp (-r/\xi_0)$, where $\xi_0$ is the initial correlation length of the system. 
Thus, the characteristic IR length scale does not change in time.
Instead, the scaling takes place in the UV giving rise to a broadening Gaussian spatial correlation function during the ordering dynamics, and thus to a sharpening Gaussian momentum-space structure factor \cite{Rutenberg1995a.PhysRevLett.74.3836}.
In contrast, the ordering process in our isolated spinor gas is driven by non-linear dynamics of the spinor field, leading to a bi-directional transport of excitations in momentum space.
This transport redistributes spin-wave excitations from an intermediate scale to both, smaller and larger wave numbers. 
Thereby, the correlation length  $L_{\Lambda}(t)\sim k_{\Lambda}(t)^{-1}$ grows in time as $L_{\Lambda} (t)\sim t^{1/4}$.
Note that a similar behavior of the correlation length has been reported in the one-dimensional $p$-state clock model for a moderate-sized $p > 4$ in Ref.~\cite{Andrenacci2006a}. 
We expect the coarsening dynamics described by this model to be closer to that of our system where the kink-like defects in the transversal spin are accompanied by phase jumps similar to the phase steps occurring in the $p$-state clock model.

Self-similar evolution within a spatio-temporal scaling regime during the phase-ordering process of a non-equilibrium system is understood to
generally occur as a transient phenomenon on the way to equilibrium. 
To study this transient nature we extract the time evolution of the characteristic momentum scales $k_{\Lambda}(t) \sim L_{\Lambda}(t)^{-1}$ and $k_{\lambda}(t) \sim L_{\lambda}(t)^{-1}$ by means of fitting the scaling form  \eq{twoLScalingFunction} to the structure factor $S(k,t)$ for evolution times up to $t = 2000 \, t_\mathrm{s}$ (see \Fig{StructureFactorUltraLong} for $S(k,t)$ at time scales beyond $t = 200 \, t_s$). 
We find that the IR scale $L_{\Lambda}(t)$ shows scaling with $\beta \simeq 0.25$ for all times considered in our simulations (see \Fig{Flowplot}a).
To retain the IR scaling, energy has to be transported to the UV. 
In the case of the bi-directional scaling evolution the energy transported to the UV leads to the sharpening of kink-like defects. 
However, defects in the spin degree of freedom are expected to have a natural minimal width on the order of the spin healing length. 
As the UV scale approaches this length scale, i.e.~as $L_\lambda(t)  = 2 \pi/ k_\lambda (t)\to 1$, we thus observe  that the UV scaling exponent starts to deviate from $\beta^{\prime} \simeq -0.17$ (see \Fig{Flowplot}b) causing the system to leave the regime of two-scale universal scaling dynamics.  
The deviation of the scaling exponent becomes clearly visible around $t \simeq 500 \, t_\mathrm{s}$.
This behavior is accompanied by a build-up of a thermal tail in the range of momenta larger than $k_{\mathrm{UV}}^>$, which instead stores the transported energy (cf.~\Fig{StructureFactorUltraLong}). 

Although the system leaves the regime of two-scale universal scaling dynamics at $t \simeq 500 \, t_\mathrm{s}$, it remains close to the non-thermal fixed point as the IR scaling exponent is unaffected for evolution times up to $t = 2000 \, t_\mathrm{s}$.
At a later point in time, which is presently beyond the reach of our simulations, we expect the rising mean kinetic energy in the thermal tail as well as the finite size of the system to induce the system to move away from the fixed point and towards final equilibrium.

In this work, we have numerically demonstrated universal self-similar dynamics in a one-dimensional ferromagnetic spin-1 Bose gas characterized by two separate time-evolving scales. 
While the IR scale increases as $L_{\Lambda}(t)\sim t^{\,\beta}$, with $\beta \simeq 0.25$, the UV scale decreases as $L_{\lambda}(t)\sim t^{\, \beta'}$, with $\beta^{\prime} \simeq -0.17$.  
Our results show that universal scaling evolution at a non-thermal fixed point is possible in a purely one-dimensional geometry, in contrast to standard arguments based on kinematic constraints prevailing for elastic collisions in 1D.
The reported scaling is amenable to experiments with ultracold Bose gases while anticipated to be relevant also for very different systems in the  relativistic realm. 

\textit{Note added.} After the completion of this paper a non-thermal fixed point associated with pair-annihilation of magnetic solitons has been reported for a one-dimensional antiferromagnetic spin-1 Bose gas  \cite{Fujimoto2018b}.

\textit{Acknowledgments.} 
We thank J.~Berges, P.B.~Blakie, I.~Chantesana,  S.~Erne, K.~Geier, S.~Heupts, M.~Karl, P.G.~Kevrekidis, P.~Kunkel, S.~Lannig, D.~Linnemann, A.N.~Mikheev, A.~Pi{\~n}eiro Orioli, J.~Schmiedmayer, H.~Strobel, and L.~Williamson for discussions and collaborations on related topics. 
This work was supported by the Horizon-2020 framework programme of the European Union (FET-Proactive, AQuS, No. 640800, ERC Advanced Grant EntangleGen, Project-ID 694561), by Deutsche Forschungsgemeinschaft (SFB 1225 ISOQUANT), by Deutscher Akademischer Austauschdienst (No.~57381316), and by Heidelberg University (CQD, HGSFP).

\clearpage
\begin{appendix}
\begin{center}
\textbf{APPENDIX}
\end{center}
\setcounter{equation}{0}
\setcounter{table}{0}
\makeatletter

\section{Initial state and post-quench dynamics}

In this appendix we briefly discuss the representation of the ground states of our model in the polar and easy-plane phases and discuss the semi-classical numerical methods with which we have obtained our results presented in the main text.

\subsection{Ground states in polar and easy-plane phase}
\label{app:SBPhaseTransition}

In the case of ferromagnetic spin interactions ($c_{1}=-1$) and for a positive quadratic Zeeman energy $q$, the equilibrium system exhibits two different phases separated by a quantum phase transition that breaks the $U(3)$ spin symmetry of the ground state 
\cite{Kawaguchi2012a.PhyRep.520.253}. For 
$q> 2$ the system is in the polar phase where the mean-field ground state, given by the state vector
\begin{equation}
\vec{\Phi}_{\mathrm{P}} = e^{i \varphi} 
\begin{pmatrix}
0 \\ 1 \\ 0
\end{pmatrix},
\end{equation}
is unmagnetized.
Here 
$\varphi$ is a global phase distinguishing different realizations of the spontaneous symmetry breaking. 

For $0 < q <  2$ 
the system is in the easy-plane ferromagnetic phase in which the mean-field ground state reads
\begin{equation}
  \vec{\Phi}_{\mathrm{EP}} = 
  \frac{e^{i \varphi}} 2 
  \begin{pmatrix}
  e^{-i \phi}\sqrt{ 1 - q/2}   \\ \sqrt{  2 + q }  \\ e^{i \phi} \sqrt{ 1 - q/2}  
  \end{pmatrix}.
\end{equation}
Here $\phi$ denotes the angle with respect to the spin-$x$-axis. 
This ground state gives rise to the mean spin vector lying in the transversal spin plane, with magnetization $ \lvert F_{\perp}\rvert =  (1- q^{2}/4 )^{1/2}$.

\begin{figure}
\includegraphics[width=0.97\columnwidth]{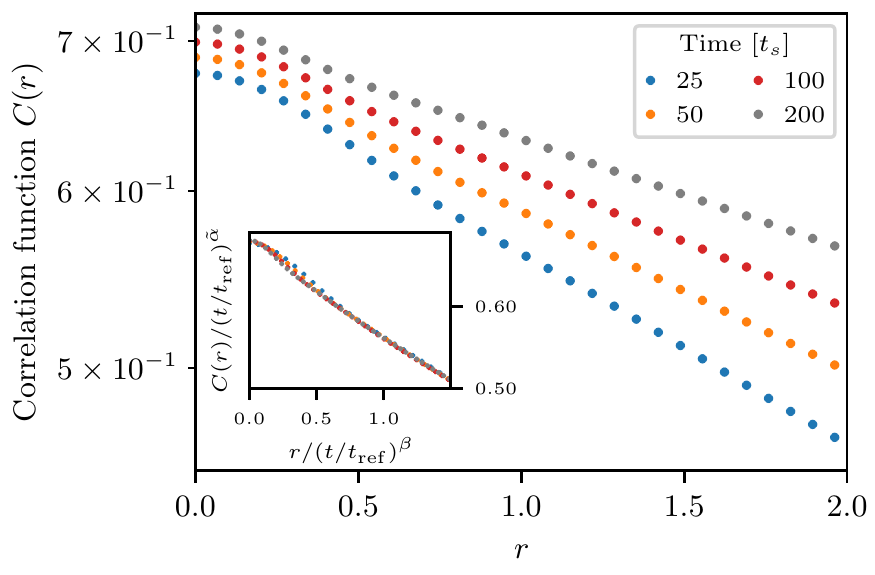}
\caption{\label{fig:CorrelationFunction1}
Spatial first-order coherence function $C(r,t)$ within the temporal scaling regime. 
$C(r,t)$ is calculated by applying a Fourier transform to the structure factor $S(k,t)$. 
At larger distances the correlation function decays exponentially $C(r,t) \sim \exp (-r/L_{\Lambda})$, with time-evolving correlation length $L_{\Lambda} (t)\sim t^{\, \beta}$.
The inset shows $C(r,t)$ rescaled with $\tilde{\alpha} = 0.02$ and $\beta = 0.25$ for reference time $t_{\mathrm{ref}} = 25 \, t_\mathrm{s}$ such that the data collapses at large distances. 
This enables to observe the shrinking of the characteristic UV length scale $L_{\lambda}(t)$ occurring at distances below $r \simeq 0.5$ . 
Note the semi-log scale of the inset.
}
\end{figure}

\subsection{Simulation methods}
\label{app:SimulationMethods}

We consider out-of-equilibrium dynamics after a sudden quench, starting from a homogeneous condensate in the $m_{F}=0$ component, $\phi_{0}(x)\equiv1$. 
We follow the time evolution by solving the coupled Gross-Pitaevskii equations (GPEs)
\begin{equation} \label{eq:GPE}
i \partial_t \vec{\Phi} =   \left ( - \frac {\partial^2}{\partial x^2} + q f_z^2 + c_0 n + c_1 \vec{F}\cdot\vec{f} \right) \vec{\Phi}
\end{equation}
by means of a spectral split-step algorithm.
We compute the time evolution of correlation functions within the semi-classical truncated Wigner approximation \cite{Blakie2008a, Polkovnikov2010a}. 

We consider experimentally relevant parameters for $^{87}$Rb in the $F=1$ hyperfine manifold, with $c_0 = 100$. 
The initial condensate density is $n_0 = 4.5 \cdot 10^4 \, \xi_\mathrm{s}^{-1}$.
The simulations are performed on a one-dimensional grid with $N_{\mathrm{g}}=4096$ grid points and periodic boundary conditions.
The corresponding physical length is $\mathcal{L}= 554\,\xi_\mathrm{s}$.

The initial state is given by a zero-temperature mean-field ground state in the polar phase, for $q_i \gg 2 $, 
with additional quantum noise sampled from the positive definite Wigner distribution of the vacuum and set into the Bogoliubov modes of the polar condensate,
\begin{equation}
\vec{\Phi} \left (x\right) = 
\begin{pmatrix}
 0 \\1 \\ 0
\end{pmatrix} + \sum_ {k}
\begin{pmatrix}
a_{k,1} e^{i k x} \\
a_{k,0} u_k e^{i k x} -  a_{k,0}^* v_k e^{-i k x} \\
a_{k,-1} e^{i k x}  \\
\end{pmatrix}.
\end{equation}
We again omit the tilde on the rescaled quantities $\tilde k = k\xi_\mathrm{s}$, $\tilde a_{\tilde k,m}=a_{k,m}/(\xi_\mathrm{s}\sqrt{n_{0}})$. 
The mode functions $a_{k,m}$ are complex Gaussian random variables with
\begin{equation}
\langle a_{k, m}^{\dagger} a_{k', m'} \rangle = \frac{1}{2} \delta_{mm'}\delta_{k,k'},
\end{equation}
which corresponds to adding an average occupation of half a particle in each mode $k$.
The Bogoliubov mode functions are given by 
\begin{equation}
u_k = \sqrt{ \frac {\epsilon_k +  c_0} {2 \sqrt{\epsilon_k \left(\epsilon_k + 2 c_0 \right)}} + \frac 1 2},    \qquad v_k = \sqrt{u_k^2 -1},
\end{equation}
with mode energy $\epsilon_k = k^2$.

Calculating observables using the truncated Wigner method requires averaging over many trajectories. 
We find that, in our one-dimensional geometry,  a sufficient convergence of the observables is reached after averaging over $\gtrsim10^{3}$ trajectories.

\begin{figure}[t]
\includegraphics[width = 0.97\columnwidth]{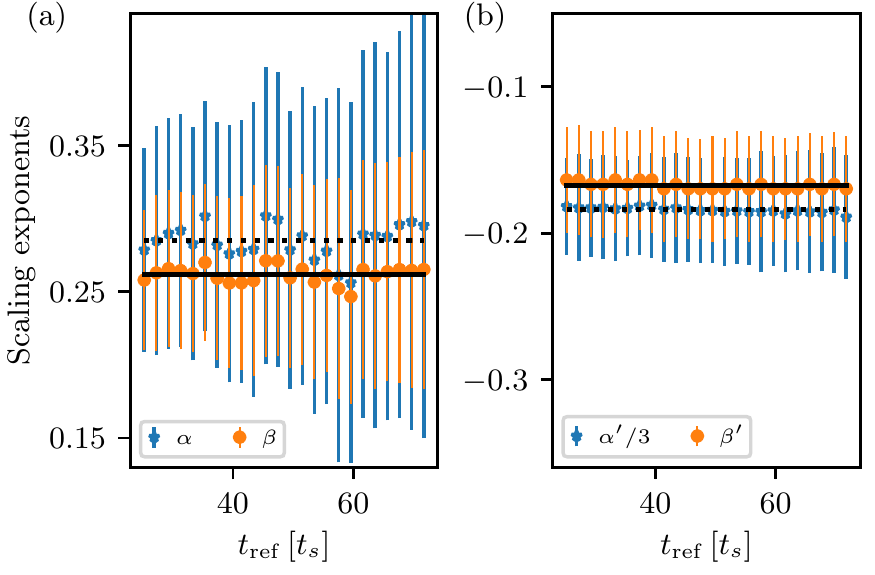}
\caption{\label{fig:ScalingExponents} 
(a) Scaling exponents $\alpha$ (blue stars) and $\beta$ (orange circles) obtained from least-square rescaling fits  to $S(k,t)$ within the time window $[t_{\mathrm{ref}}, t_{\mathrm{ref}} + \Delta t]$ with $\Delta t = 120 \, t_\mathrm{s}$. The scaling exponents are independent of the reference time $t_{\mathrm{ref}}$.
A constant fit to the data reveals
$\alpha = 0.284 \pm 0.013$ (dashed line) and $\beta = 0.261\pm 0.008$ (solid line). 
The error is given by the standard deviation of all data points as they are not statistically independent.
(b) Scaling exponents $\alpha^{\prime}/3$  (blue stars) and $\beta^{\prime}$  (orange circles) obtained by the same method and within the same time window as in (a). 
 A constant fit to the data yields
$\alpha^{\prime}/3 = -0.183 \pm 0.002$ (dashed line) and $\beta^{\prime}= -0.168\pm 0.004$ (solid line). Error computed as in (a). 
}
\end{figure}

\section{Universal scaling dynamics}
\label{app:Correlationlength}

In this appendix, we discuss, in more detail, the position-space correlation function in the scaling regime and demonstrate how the system departs from scaling during the late period of the evolution.

\subsection{Spatial correlation function}

The two different characteristic length scales that undergo universal scaling dynamics in the system can also be studied by means of the position-space correlation function $C(r,t)$ which is calculated by applying a  Fourier transform to the structure factor $S(k,t)$. Fig.~\ref{fig:CorrelationFunction1} shows the position-space correlation function for distances $0 \leq r  \leq 2$ within the temporal scaling regime.
The shrinking of the characteristic UV length scale $L_{\lambda} (t) \sim t^{\, \beta^{ \prime}} $ is found below distances $r \simeq 0.5$. 
It is related to the quadratic part of the correlation function at short distances becoming steeper (see inset of Fig.~\ref{fig:CorrelationFunction1}).
Note that the effect is small due to the slow scaling with $\beta^{\prime} \simeq -0.17$.
At larger distances the correlation function is given by
\begin{equation}
C(r,t) \sim \exp \left(- \frac r {L_{\Lambda}}\right) ,
\end{equation}
with time-evolving correlation length $L_{\Lambda}(t) \sim t^{\, \beta}$. 
The growth of the correlation length in time is associated with the decrease of the slope of the correlation function drawn in semi-logarithmic representation.
This behavior is in contrast to coarsening dynamics of the 1D XY model where the Gaussian short-distance part grows in space while the slope of the exponential tail remains constant.

\subsection{Scaling regime}
\label{app:Reference time}

Within the spatio-temporal scaling period, the scaling exponents are found to be independent of the reference time $t_{\text{ref}}$. 
Performing the least-square rescaling analysis for different reference times $t_\mathrm{ref}$,  within the constant time window $[t_{\mathrm{ref}}, t_{\mathrm{ref}} + \Delta t]$ with $\Delta t = 120 \, t_\mathrm{s}$, we find that the scaling exponents settle to a constant value at $t_{\text{ref}} \simeq 25\, t_\mathrm{s}$ which marks the onset of the scaling regime (see \Fig{ScalingExponents}).

A constant fit to the extracted IR scaling exponents for $25\, t_\mathrm{s} \leq t_{\mathrm{ref}} \leq 74\, t_\mathrm{s}$ yields 
$\alpha = 0.284\pm 0.013$ and $\beta = 0.261\pm 0.008 \,$.
For the UV scaling exponents we find $\alpha^{\prime}= -0.549 \pm 0.006$ and $\beta^{\prime}= -0.168\pm 0.004$.
The error is given by the standard deviation of all data points, which are not statistically independent. 
Making use of analyzing the scaling exponents for various reference times incorporates fluctuations of the scaling exponents caused by statistical errors in each reference spectrum thus leading to a more accurate determination of the universal exponents.

\begin{figure}[t]
\includegraphics[width=0.96\columnwidth]{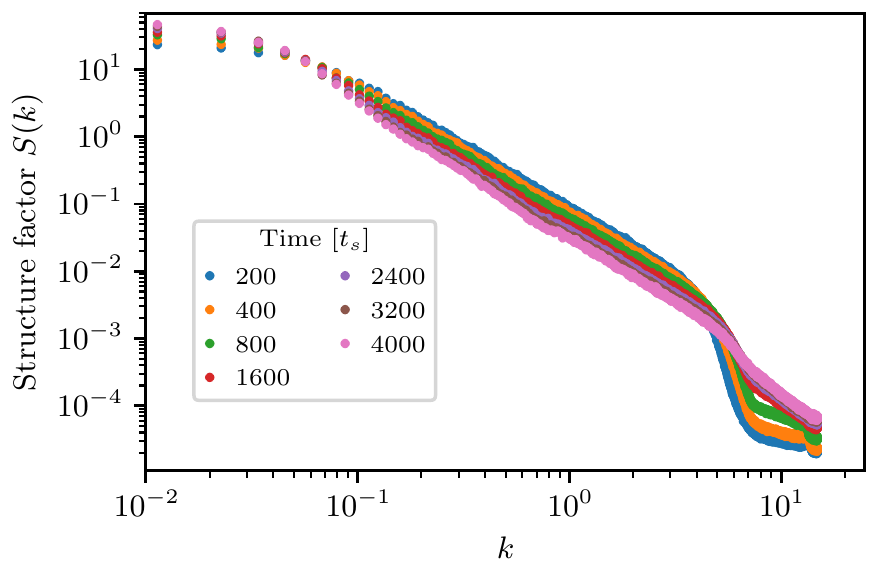}
\caption{\label{fig:StructureFactorUltraLong}
Time evolution of the structure factor $S(k)$ for $q_\mathrm{f}= 0.9$ on time scales beyond the scaling regime presented in the main text. 
The data indicates that bi-directional self-similar dynamics can be observed even up to $t \simeq 400\, t_\mathrm{s}$ as the scaling function remains the same. 
For $t \geq 800\, t_\mathrm{s}$ a thermal tail given by an approximate $k^{-2}$ power law is present in the UV. 
The mean kinetic energy in the tail gradually increases as time evolves up to $t = 4000\, t_\mathrm{s}$. 
}
\end{figure}

\subsection{Departure from scaling}
\label{app:Beyondscaling}

Scaling dynamics during the phase ordering process of a non-equilibrium system generically represents a transient process on the way to equilibration. 
Therefore we expect the system to leave the scaling regime at some time and relax back to its equilibrium state. 

We are able to observe indications of this behavior when simulating the system $20$ times longer than presented in the main text. 
\Fig{StructureFactorUltraLong} shows the time evolution of the structure factor $S(k,t)$ for $q_\mathrm{f}= 0.9$ on time scales up to $t= 4000\, t_\mathrm{s}$. 
As shown in \Fig{Flowplot}, the system leaves the two-scale self-similar regime at $t \simeq 500 \, t_\mathrm{s}$ consistent with the data depicted in \Fig{StructureFactorUltraLong}.
For $t \geq 500\, t_\mathrm{s}$ a thermal tail given by an approximate $k^{-2}$ power law is formed in the UV.
The temperature of the state is characterized by the slope of the power law. 
It slowly increases as time evolves up to $t = 4000\, t_\mathrm{s}$. 
We expect the rising mean kinetic energy in the thermal tail to eventually cause the breakdown of the IR scaling. 
In consequence, the system is driven away from the non-thermal fixed point towards equilibrium.
Note that the final equilibration process is beyond the time scales considered in our numerical simulations and depends on the particular IR and UV boundary conditions realized in the setup.

\end{appendix}


%

\end{document}